\documentclass[conference,compsoc,onecolumn]{IEEEtran}
\ifCLASSOPTIONcompsoc
  \usepackage[nocompress]{cite}
\else
  \usepackage{cite}
\fi

\ifCLASSINFOpdf
\else
\fi
\usepackage{amsmath}

\usepackage{array}
\ifCLASSOPTIONcompsoc
\usepackage[caption=false,font=footnotesize,labelfont=sf,textfont=sf]{subfig}
\else
\usepackage[caption=false,font=footnotesize]{subfig}
\fi
\usepackage{url}
\usepackage{graphicx}

\usepackage{algorithm,algpseudocode}
\algrenewcommand\algorithmicrequire{\textbf{Input:}}
\algrenewcommand\algorithmicensure{\textbf{Output:}}
\usepackage{amsmath,amsfonts,amssymb,amsthm}
\usepackage{xcolor}
\definecolor{ocre}{RGB}{0,173,239}

\usepackage{tikz}
\usepackage{pgfplots}
\pgfplotsset{compat=1.18}

\usetikzlibrary{arrows,matrix,positioning,fit}
\usetikzlibrary{patterns,decorations.pathreplacing}
\tikzset{%
  highlight/.style={rectangle,rounded corners,fill=ocre!50,draw=ocre!80,
     fill opacity=0.2,inner sep=0pt,text opacity=1}
}

\usepackage{graphicx} 
\usepackage{float}
\usepackage[all]{xy}
\usepackage{tikz-cd}
\usepackage[shortlabels]{enumitem}
\usepackage{dsfont}
\usepackage{tikz}
\usepackage{pbox}
\usetikzlibrary{backgrounds}
\pgfdeclarelayer{background2}
\pgfdeclarelayer{background}
\pgfdeclarelayer{foreground}
\pgfdeclarelayer{foreground2}
\pgfsetlayers{background2,background,main,foreground,foreground2}

\makeatletter
\renewcommand*\env@matrix[1][*\c@MaxMatrixCols c]{%
  \hskip -\arraycolsep
  \let\@ifnextchar\new@ifnextchar
  \array{#1}}
\makeatother

\usepackage[edges]{forest}
\usepackage{array}
\newcolumntype{C}[1]{>{\centering\arraybackslash}p{#1}}
\newcolumntype{L}[1]{>{\raggedright\arraybackslash}p{#1}}
\usetikzlibrary{arrows.meta,shadows}
\usetikzlibrary{trees,positioning,shapes,shadows,arrows}

\usepackage[hypertexnames=false]{hyperref}
\hypersetup{pdfborder=0 0 0}

\algnewcommand{\LineComment}[1]{\State \(\triangleright\) #1}

\usepackage{tablefootnote} % for table footnotes

\usepackage{lastpage} % for page out of
\usepackage{fancyhdr} % for page out of
\newlength{\gpgpuElemSep}
\newlength{\gpgpuElemSize}
\usepackage{listings}
\usepackage{caption}
\makeatletter
\let\old@lstKV@SwitchCases\lstKV@SwitchCases
\def\lstKV@SwitchCases#1#2#3{}
\makeatother
\usepackage{lstlinebgrd}
\makeatletter
\let\lstKV@SwitchCases\old@lstKV@SwitchCases

\lst@Key{numbers}{none}{%
    \def\lst@PlaceNumber{\lst@linebgrd}%
    \lstKV@SwitchCases{#1}%
    {none:\\%
     left:\def\lst@PlaceNumber{\llap{\normalfont
                \lst@numberstyle{\thelstnumber}\kern\lst@numbersep}\lst@linebgrd}\\%
     right:\def\lst@PlaceNumber{\rlap{\normalfont
                \kern\linewidth \kern\lst@numbersep
                \lst@numberstyle{\thelstnumber}}\lst@linebgrd}%
    }{\PackageError{Listings}{Numbers #1 unknown}\@ehc}}
\makeatother

\def\FunctionD(#1){2*(#1)-1}%
\def\FunctionM(#1,#2){((3*(#1)^2)*10^9)/(#2)}
\newcommand{\PreserveBackslash}[1]{\let\temp=\\#1\let\\=\temp}
\newcommand\hlightcolor[1]{\tikz[overlay, remember picture,baseline=-\the\dimexpr\fontdimen22\textfont2\relax]\node[rectangle,fill=magenta!50,rounded corners,fill opacity=0.2,draw=magenta!80,text opacity=1] {$#1$};} 
\usepackage{soul} %strikethrough font
\setstcolor{blue}
\usepackage{accents}

\definecolor{myblue4} {RGB}{66,129,196}
\tikzstyle{myline} = [very thick, draw=myblue4, fill=myblue4, shape=rectangle, inner sep=10pt, inner ysep=20pt]
\tikzstyle{mybox} = [very thick, draw=myblue4, fill=white, shape=rectangle, inner sep=10pt, inner ysep=20pt]

\tikzstyle{mybox1} = [very thick, draw=white, fill=white, shape=rectangle, inner sep=10pt, inner ysep=20pt]

\usepackage{adjustbox}
\usepackage{fourier-orns}
\usepackage{multirow}
\usepackage{tikz}
\usepackage{pgfplotstable}
\usepackage{ifthen}

\DeclareCaptionFormat{algor}{%
\hrulefill\par\offinterlineskip\vskip1pt%
\textbf{#1#2}#3\offinterlineskip\hrulefill}
\DeclareCaptionStyle{algori}{singlelinecheck=off,format=algor,labelsep=space}
\usepackage{bm}
\usepackage{setspace}

\usepackage{multicol}
\usepackage{pdflscape}

\usepackage{longtable}
\usepackage{stackengine}

\usepackage{subcaption}
\pgfplotscreateplotcyclelist{my_black_white}{%
solid, every mark/.append style={solid, fill=gray}, mark=*\\%
solid, every mark/.append style={solid, fill=gray}, mark=square*\\%
solid, every mark/.append style={solid, fill=gray}, mark=star\\%
solid, every mark/.append style={solid, fill=gray}, mark=diamond*\\%
solid, every mark/.append style={solid, fill=gray}, mark=otimes\\%
solid, every mark/.append style={solid, fill=gray}, mark=Mercedes star\\%
solid, every mark/.append style={solid, fill=gray}, mark=triangle*\\%
solid, every mark/.append style={solid, fill=gray}, mark=oplus\\%%%%%
densely dotted, every mark/.append style={solid, fill=gray}, mark=*\\%
densely dotted, every mark/.append style={solid, fill=gray}, mark=square*\\%
densely dotted, every mark/.append style={solid, fill=gray}, mark=star\\%
densely dotted, every mark/.append style={solid, fill=gray}, mark=diamond*\\%
densely dotted, every mark/.append style={solid, fill=gray}, mark=otimes\\%
densely dotted, every mark/.append style={solid, fill=gray}, mark=Mercedes star\\%
densely dotted, every mark/.append style={solid, fill=gray}, mark=triangle*\\%
densely dotted, every mark/.append style={solid, fill=gray}, mark=oplus\\%%%%%
dashed, every mark/.append style={solid, fill=gray}, mark=*\\%
dashed, every mark/.append style={solid, fill=gray}, mark=square*\\%
dashed, every mark/.append style={solid, fill=gray}, mark=star\\%
dashed, every mark/.append style={solid, fill=gray}, mark=diamond*\\%
dashed, every mark/.append style={solid, fill=gray}, mark=otimes\\%
dashed, every mark/.append style={solid, fill=gray}, mark=Mercedes star\\%
dashed, every mark/.append style={solid, fill=gray}, mark=triangle*\\%
dashed, every mark/.append style={solid, fill=gray}, mark=oplus\\%%%%%
dashdotted, every mark/.append style={solid, fill=gray}, mark=o\\%
dashdotted, every mark/.append style={solid, fill=white}, mark=square*\\%
}

\usepackage{subcaption}
\begin{document}
\pagenumbering{arabic}
\pagestyle{plain}

\title{ML-Based Optimum Sub-system Size\\ for the GPU Implementation of the Tridiagonal Partition Method}

\author{
\IEEEauthorblockN{Milena Veneva\IEEEauthorrefmark{1}, \emph{milena.p.veneva@gmail.com}}
\IEEEauthorblockA{\IEEEauthorrefmark{1}RIKEN Center for Computational Science, R-CCS, 7-1-26 Minatojima-minami-machi, Chuo-ku, \\Kobe, Hyogo 650-0047, Japan}
} 

% make the title area
\maketitle
\thispagestyle{plain}

\section*{Abstract}
This paper presents a machine learning~(ML)-based heuristic for finding the optimum sub-system size for the CUDA implementation of the parallel partition algorithm. 
Computational experiments for different system of linear algebraic equation (SLAE) sizes are conducted, and the optimum sub-system size for each of them is found empirically.
To estimate a model for the sub-system size, we perform the k-nearest neighbors~(kNN) classification method. Statistical analysis of the results is done. By comparing the predicted values with the actual data, the algorithm is deemed to be acceptably good. Next, the heuristic is expanded to work for the recursive parallel partition algorithm as well. An algorithm for determining the optimum sub-system size for each recursive step is formulated. A kNN model for predicting the optimum number of recursive steps for a particular SLAE size is built.

%%%%%%%%%%%%%%%%%%%%%%%%%%%%%%%%%%%%%%%%%%%%%%%%%%%%%%%%%%%%%%%%%%%%%%%
\section{Introduction} 
\label{intro}

The parallel partition algorithm for solving systems of linear algebraic equations~(SLAEs) 
suggested in~\cite{Austin:2004} is an efficient numerical technique for solving SLAEs with tridiagonal
coefficient matrices. It works by splitting the original matrix into smaller sub-matrices and solving the corresponding smaller SLAEs in parallel.
Originally designed for use with a large number of processors, this algorithm was implemented using MPI (Message Passing Interface) technology in~\cite{Austin:2004}. 

The algorithm consists of three stages. In the first one, we partition the initial SLAE into sub-systems, and for each of them compute two interface equations. In the second Stage, the interface equations are assembled into an interface system which is also tridiagonal (and diagonally dominant if the initial one is such~\cite{Austin:2004}). After we find the solution of the interface system, we have found the values of some of the unknowns of the initial system. Therefore, we can substitute them and solve the rest of the sub-systems, which are now independent, in parallel. This puts an end to Stage 3. In our CUDA~\cite{cuda} implementation, the first Stage is run on the device. The second Stage is done on the host and therefore requires device-to-host~(D2H) memory transfer after Stage 1 and host-to-device~(H2D) memory transfer after Stage 2. Finally, Stage 3 is also run on the device. The parallel partition method includes a non-recursive variant, which will be examined in Section~\ref{sec:non_rec}, and a recursive variant, discussed in Section~\ref{sec:rec}.

The development of high performance computing~(HPC) applications typically consists of two key phases: writing code that functions correctly and then optimizing that code to enhance performance. We present one of the optimizations made to our CUDA implementation, namely building a heuristic for finding the optimum sub-system size by using tools frequently used in the modern artificial intelligence~(AI)-focused approaches like the k-nearest neighbors classification algorithm.

The nature of the parallel partition method is such that the initial SLAE with $N$ unknowns is partitioned into a number of sub-systems with $m$ unknowns each.
The size of the SLAE $N$ that the user solves is usually determined by the problem they need to solve, while the size of the sub-system within the parallel partition method $m$ is a parameter that needs to be tuned. Thus, having a heuristic which predicts the optimum sub-system size for each SLAE size would allow us to gain additional GPU performance.

%%%%%%%%%%%%%%%%%%%%%%%%%%%%%%%%%%%%%%%%%%%%%%%%%%%%%%%%%%%%%%%%%%%%%%%
\section{Building a heuristic for the optimum sub-system size}
\label{sec:non_rec}

The heuristic processes of optimization in solving SLAEs on GPUs are summarized as
of NVIDIA GPU RTX 2080~Ti~\cite{2080a,2080b} for SLAE sizes $10^{i}$, 
$2\times10^{i}$, $4\times10^{i}$, $5\times10^{i}$, $8\times10^{i}$, $i=2,3,\dots,7$, and $4.5\times10^{3}$, $2.5\times10^{4}$, $3\times10^{4}$, $6\times10^{4}$, $7\times10^{4}$, $7.5\times10^{4}$, $10^{8}$. Between 11 and 18 different sub-system sizes in the interval $[4;1250]$ were tested for each of the 37 considered SLAE sizes. The rest of the parameters are: 256 CUDA threads per block, the number of CUDA streams was determined according to the optimum stream heuristic~\cite{Veneva_2024} and FP64 precision. The average time of several runs was taken. The results are summarized in Table~\ref{tab:my_label}.

\begin{table}[!htbp]
\caption{Observations on the optimum sub-system size. The 1$^{\textrm{st}}$ column shows the size of the initial SLAE $N$, the 2$^{\textrm{nd}}$ -- the experimentally found optimum sub-system size $m$, the 3$^{\textrm{rd}}$ is the number of CUDA streams that were used (based on~\cite{Veneva_2024}), the 4$^{\textrm{th}}$ is the computational time for the partition method when using the opt $m$, the other three columns are explained below.}
\label{tab:my_label}
\centering
\resizebox{0.8\textwidth}{!}{%
\begin{tabular}{|r|r|r|r|r|r|r|}
\hline
$N$ & opt $m$ & \#streams & time opt & corrected & time corrected & difference \\
&&& $m$ [ms] (1) & opt $m$ & opt $m$ [ms] (2) & (1) - (2) \\\hline\hline
$10^{2}$ & 4 & 1 & 0.310275 & 4 & -- & --\\
$2\times10^{2}$ & 4 & 1 & 0.315868 & 4 & -- & --\\
$4\times10^{2}$ & 4 & 1 & 0.327477 & 4 & -- & --\\
$5\times10^{2}$ & 4 & 1 & 0.325367 & 4 & -- & --\\
$8\times10^{2}$ & 4 & 1 & 0.340679 & 4 & -- & --\\
$10^{3}$ & 4 & 1 & 0.331446 & 4 & -- & --\\
$2\times10^{3}$ & 4 & 1 & 0.351094 & 4 & -- & --\\
$4\times10^{3}$ & 4 & 1 & 0.373837 & 4 & -- & --\\
$4.5\times10^{3}$ & 4 & 1 & 0.385070 & 4 & -- & --\\\hline
$5\times10^{3}$ & 8 & 1 & 0.380488 & 8 & -- & --\\
$8\times10^{3}$ & 8 & 1 & 0.424161 & 8 & -- & --\\
$10^{4}$ & 8 & 1 & 0.438337 & 8 & -- & --\\
$2\times10^{4}$ & 8 & 1 & 0.536961 & 8 & -- & --\\
$2.5\times10^{4}$ & 8 & 1 & 0.591000 & 8 & -- & --\\\hline
$3\times10^{4}$ & 16 & 1 & 0.614149 & 16 & -- & --\\
$4\times10^{4}$ & 16 & 1 & 0.711075 & 16 & -- & --\\
$5\times10^{4}$ & 16 & 1 & 0.785274 & 16 & -- & --\\\hline
$6\times10^{4}$ & 20 & 1 & 0.874056 & 20 & -- & --\\
$7\times10^{4}$ & 35 & 1 & 0.956710 & 20 & 0.957520 & -0.000810\hspace{0em} \\
$7.5\times10^{4}$ & 40 & 1 & 0.995135 & 20 & 1.002325 & -0.00719\hspace{0.5em} \\\hline
$8\times10^{4}$ & 32 & 1 & 1.034019 & 32 & -- & --\\
$10^{5}$ & 40 & 1 & 1.195640 & 32 & 1.196261 & -0.000621\hspace{0em} \\
$2\times10^{5}$ & 64 & 2 & 1.857711 & 32 & 1.931349 & \textbf{-0.073638}\\
$4\times10^{5}$ & 64 & 4 & 3.270235 & 32 & 3.339023 & \textbf{-0.068788}\\
$5\times10^{5}$ & 40 & 8 & 4.043336 & 32 & 4.089002 & -0.045666\hspace{0em} \\
$8\times10^{5}$ & 64 & 8 & 6.055748 & 32 & 6.237866 & \textbf{-0.182118}\\
$10^{6}$ & 32 & 8 & 7.635039 & 32 & -- & --\\
$2\times10^{6}$ & 32 & 16 & 14.49496\hspace{0.5em} & 32 & -- & --\\
$4\times10^{6}$ & 32 & 32 & 27.83609\hspace{0.5em} & 32 & -- & --\\
$5\times10^{6}$ & 32 & 32 & 34.51819\hspace{0.5em} & 32 & -- & --\\
$8\times10^{6}$ & 64 & 32 & 53.92044\hspace{0.5em} & 32 & 54.36878 & -0.44834\hspace{0.5em} \\
$10^{7}$ & 32 & 32 & 66.71282\hspace{0.5em} & 32 & -- & --\\\hline
$2\times10^{7}$ & 64 & 32 & 131.0139\hspace{1em} & 64 & -- & --\\
$4\times10^{7}$ & 64 & 32 & 259.8288\hspace{1em} & 64 & -- & --\\
$5\times10^{7}$ & 64 & 32 & 323.7364\hspace{1em} & 64 & -- & --\\
$8\times10^{7}$ & 64 & 32 & 516.1501\hspace{1em} & 64 & -- & --\\
$10^{8}$ & 64 & 32 & 643.1100\hspace{1em} & 64 & -- & --\\
\hline\hline
\end{tabular}}
\end{table}

\subsection{Considerations}

Before building the heuristic for the optimum sub-system size, we start with some considerations.
\subsubsection{Optimum blockSize}
When optimizing the blockSize, which is the number of threads within a block, we should choose it in such a way that there are enough warps
in the block so as to keep the streaming multiprocessor~(SM) busy while one warp is waiting for resources, e.\,g., loading data from memory.
According to~\cite{best_pr} a programmer who strives for performance should start their experiments with between 128 and 256 threads within a CUDA block.
In the case of the CUDA implementation of the parallel partition method, choosing the blockSize to be 256 gives the best performance on NVIDIA RTX 2080~Ti GPU card. On RTX A5000~\cite{a5000a,a5000b} GPU card the smallest computational time was achieved when using blockSize of 128, but the difference between the computational times for 256 and 128 threads was negligible. blockSize smaller than 128 is not enough to supply the SM with enough active warps for the scheduler to choose between and hence to hide the latency. blockSize bigger than 256 requires too much resources, e.\,g., registers and shared memory, hence we start over-utilizing the GPU, and therefore the performance starts deteriorating. Thus, we fix the blockSize to 256.

\subsubsection{Balance between metrics}
To gain the uttermost of the GPU performance, we should balance between (1) latency hiding, (2) resource utilization, refraining from under- or over-utilization of the GPU, and (3) occupancy, which dictates how many active warps we can have.

\subsubsection{gridSize}
The data decomposition of the CUDA implementation of the partition method is such that every sub-system is processed by a single CUDA thread. Therefore, having fixed the SLAE size, and the blockSize, by optimizing the sub-system size, we dictate how many threads should run our kernels, which is the gridSize. Hence, within the partition method the gridSize depends on the sub-system size.

\subsection{Overview of existing tuning heuristics}

In the literature, there are a few different tuning approaches. Some use exhaustive search over all possible combinations of parameters, e.\,g., QUDA~\cite{quda_a,quda_b}, but this requires additional runs of the application and is energy-costly. However, such an approach would be beneficial for applications which are being run many times with the same parameters in the course of computation. 
Some other approaches promote a performance characteristic with the expectation that in the worst-case scenario, this approach is not going to lead to pathologically bad performance. Such an example comes from Thrust~\cite{thrust,calc}, which is the CUDA-based equivalent of the standard template library (STL), where the promoted characteristic is occupancy. Others use a two-step approach to gather information about the GPU limitations and the kernel that needs to be launched~\cite{Jeshani_2023}.
And others use machine learning techniques to predict combinations of parameters that would lead to optimal or near-optimal performance~\cite{Sato_2010}.

%%%%%%%%%%%%%%%%%%%%%%%%%%%%%%%%%%%%%%%%%%%%%%%%%%%%%%%%%%%%%%%%%%%%%%%%%%%

\subsection{Occupancy}

The achieved and the theoretical occupancy for the kernels responsible for Stage 1 and Stage 3 are shown in Figure~\ref{fig:occupancy_no_wd}. 
These results were achieved for the optimum sub-system size which leads to the smallest computational time for each of the listed SLAE sizes. 
The theoretical occupancy for the two kernels coincides, and therefore the reader sees only one line at level 100\%. 
Looking at the Figure, we can conclude that for SLAE size up to $4\times10^{7}$ the achieved occupancy for the kernels is smaller than $50\%$ while the theoretical occupancy is 100\%. Thus, the occupancy cannot be our main reference point if we want to minimize the computational time, that is, we cannot follow the example of Thrust.
\pgfplotsset{
    /pgfplots/my legend/.style={
        legend image code/.code={
        \draw[thick,black](-0.05cm,0cm) -- (0.3cm,0cm);%
        }
    },
    %cycle list/Set1-9,
        /pgfplots/ybar legend/.style={
        /pgfplots/legend image code/.code={%
        \draw[##1,/tikz/.cd,yshift=-0.25em]
        (0cm,0cm) rectangle (3pt,0.8em);},
   },
}

\pgfplotstableread[col sep=comma,header=true]{
A,Stage 1,Stage 3, Theoretical for 1, Theoretical for 3
$10^2$,14.57,9.34,100,100
$2\times10^2$,17.15,11.42,100,100
$4\times10^2$,24.21,15.26,100,100
$5\times10^2$,24.18,15.33,100,100
$8\times10^2$,21.80,21.84 ,100,100
$10^3$,23.99,23.97,100,100
$2\times10^3$,23.95,23.71,100,100
$4\times10^3$,23.88,23.88,100,100
$5\times10^3$,24.22,21.67,100,100
$8\times10^3$,24.23,24.14,100,100
$10^4$,24.32,23.80,100,100
$2\times10^4$,23.91,23.79,100,100
$4\times10^4$,24.13,24.19,100,100
$5\times10^4$,24.13,23.83,100,100
$8\times10^4$,24.32,24.38,100,100
$10^5$,24.50,24.35,100,100
$2\times10^5$,24.48,24.09,100,100
$4\times10^5$,24.30,23.72,100,100
$5\times10^5$,24.73,24.64,100,100
$8\times10^5$,24.75,24.79,100,100
$10^6$,24.60,24.22,100,100
$2\times10^6$,24.48,24.11,100,100
$4\times10^6$,24.36,24.32,100,100
$5\times10^6$,24.47,24.40,100,100
$8\times10^6$,24.52,24.45,100,100
$10^7$,24.63,24.54,100,100
$2\times10^7$,29.14,24.60,100,100
$4\times10^7$,52.86,29.30,100,100
$5\times10^7$,62.25,37.40,100,100
$8\times10^7$,86.11,57.26,100,100
$10^8$,88.59,68.82,100,100
}\datatable
\begin{figure}[!htb]
\centering
\resizebox{0.75\textwidth}{!}{%
\begin{tikzpicture}
\begin{axis}[
%cycle list/Set1-9,
%title={Comparison between the achieved and the theoretical occupancy for \_NO\_WD.},
enlarge y limits={value=.1,upper},
ymajorgrids = true,
major grid style={draw=white},
ymin=0.01, ymax=100,
axis x line*=bottom,
axis y line*=right,
height=10cm, width=15.5cm,
%xmode=log,
%ymode=log, 
ylabel = {Occupancy [\%]},
x tick label style={rotate=90, anchor=east},
xtick={0,...,30},
xticklabels from table={\datatable}{A},
legend cell align={left},
legend style={
    at={(1,0.95)},
    anchor=south east,
    legend columns=2,
    /tikz/every even column/.append style={column sep=0.5cm}
},
every axis plot post/.append style={
},
]

\addplot[mark=o,green!50!white,line width=1pt] table [x expr=\coordindex, y=Stage 1] {\datatable};
\addplot[mark=square,violet!80!white,line width=1pt] table [x expr=\coordindex, y=Stage 3] {\datatable};
\addplot[mark=*,blue!80!white,line width=1pt] table [x expr=\coordindex, y=Theoretical for 1] {\datatable};
\addplot[mark=*,brown!80!white,line width=1pt] table [x expr=\coordindex, y=Theoretical for 3] {\datatable};
\legend{Achieved occupancy for Stage 1,Achieved occupancy for Stage 3, Theoretical occupancy for Stage 1, Theoretical occupancy for Stage 3}
\end{axis}
\end{tikzpicture}
}
\caption{Comparison between the achieved and the theoretical occupancy.}
\label{fig:occupancy_no_wd}
\end{figure}
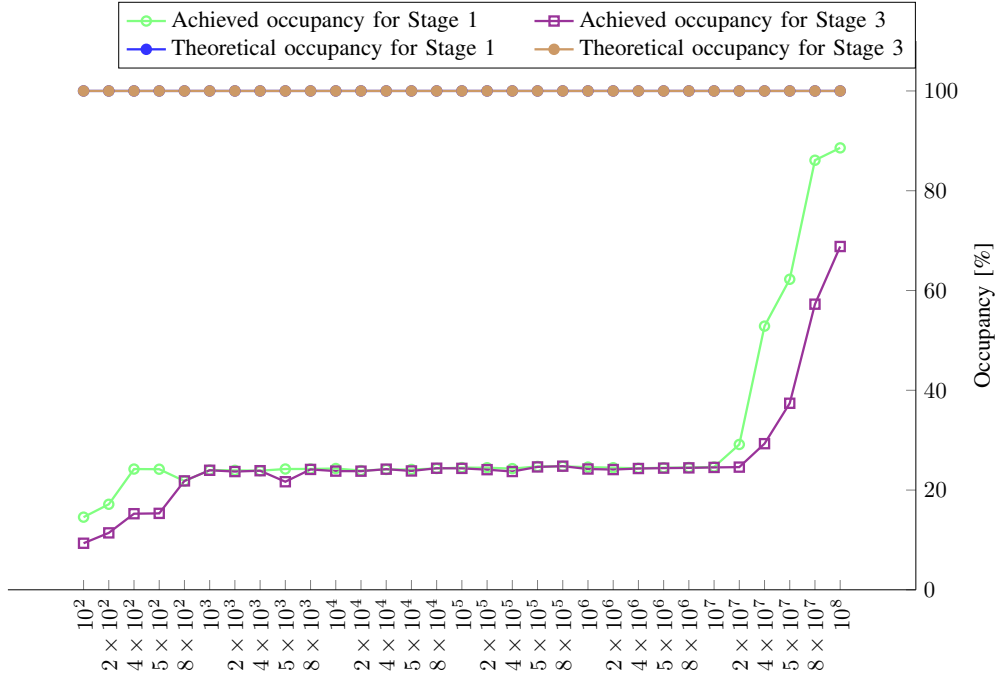

\subsection{Observations on the optimum sub-system size}
Looking at the summarized results from the computational experiment in Table~\ref{tab:my_label}, one can notice that there is a certain trend for the optimum sub-system size $m$ depending on the SLAE size, that is, we can split our data into sub-intervals, as follows:
\begin{itemize}
\item 
for $N\leq 4.5\times10^{3}$, $m = 4$; 
\item 
$4.5\times10^{3} < $ $N$ $\leq 2.5\times10^{4}$, $m = 8$; 
\item 
$2.5\times10^{4} < $ $N$ $\leq 5\times10^{4}$, $m = 16$; 
\item 
$5\times10^{4} < $ $N$ $\leq 7.5\times10^{4}$, $m = 20$;
\item 
$7.5\times10^{4} < $ $N$ $\leq 10^{7}$, $m = 32$;
\item 
$10^{7} < $ $N$ $\leq 10^{8}$, $m = 64$.
\end{itemize}
These observations are going to help us build the heuristic for the optimum sub-system size.

\subsection{Building a mathematical model for the optimum sub-system size}

It is good to have a model for building a heuristic for the optimum sub-system size $m$, not just the insight we gained in the previous Subsection.
To estimate a model for the sub-system size, we perform the k-nearest neighbors~(kNN) classification~\cite{Fix_1989, Korstanje, Banerjee_2020}. The kNN algorithm is a non-linear supervised machine learning model. The algorithm is non-parametric, meaning that there is no assumption for the data distribution. The algorithm finds the dependent~(target) variable for new data by finding $k$ nearest neighbors of the new data point, and taking the mode, i.\,e., the value that occurs most often, of the values of their dependent variables as the prediction for the dependent variable of the new data point.  

The independent variable in our case is the SLAE size, while the dependent variable is the optimum sub-system size. 
We split all the data we have for the sub-system size into two data sets -- training (neighbors), and test (for which we make predictions) -- using the Python \texttt{scikit-learn}~\cite{scikit} routine \texttt{train\_test\_split} with shuffle option turned on, and splitting ratio $3 : 1$. 
We used the \texttt{scikit-learn}~\cite{scikit} tool \texttt{GridSearchCV} to look for the best hyper-parameter $k$, which should be between 1 and the number of unique sub-system sizes. 
We applied two different approaches when fitting the data: (1) using the experimentally found optimum sub-system sizes, which are shown in the second column of Table~\ref{tab:my_label}, and (2) using the 
corrected sub-system sizes, which are shown in the fifth column of Table~\ref{tab:my_label}, and which take into account the trend we have noticed. Let us justify the shift from the experimentally found data to the corrected one. The real data showed certain fluctuations in the optimum $m$, e.\,g., take a look at the results in the intervals [$6\times10^{4}$; $7.5\times10^{4}$] and [$8\times10^{4}$; $10^{7}$]. 
%We cannot expect to be able to build a kNN model which can learn correctly from such data. 
On the other hand, as listed in the previous Subsection, there is a trend in the optimum $m$. Comparison between the computational times for the partition method when using the optimum $m$ and the corrected optimum $m$ is done in the last column of Table~\ref{tab:my_label}. In fact, in the 8 out of 37 cases when we had to make a correction, the corrected optimum $m$ came from the sub-system size that led to the second (4 cases), third (3 cases) or fourth (1 case) best computational time, and the difference between these times is relatively small as a percentage of the computational time, that is, $\leq1\%$ in 5 of the cases, and $\leq3\%$ in the other 3 cases (given in bold in Table~\ref{tab:my_label}).

One should have in mind that it was important to split and shuffle the data in such a way that the model has all possible sub-system sizes values in the training set. Otherwise, the model does not learn correctly. 
$k$ was found to be equal to 1, which is a special case of kNN also known as the nearest neighbor interpolation. This result is quite logical because it makes sense to assign as optimum sub-system size for a particular SLAE size the sub-system size of the closest SLAE size.

\begin{figure}[htb]
\centering
\subfloat[Results from the kNN classification model for the optimum sub-system size using the corrected data for $m$.]{{\fbox{\includegraphics[width=0.45\textwidth]{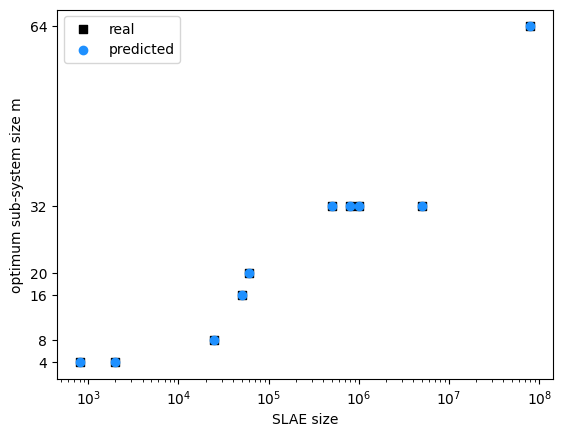}}}}
\qquad\subfloat[Results from the kNN classification model for the optimum sub-system size using the observed data for $m$.]{{\fbox{\includegraphics[width=0.45\textwidth]{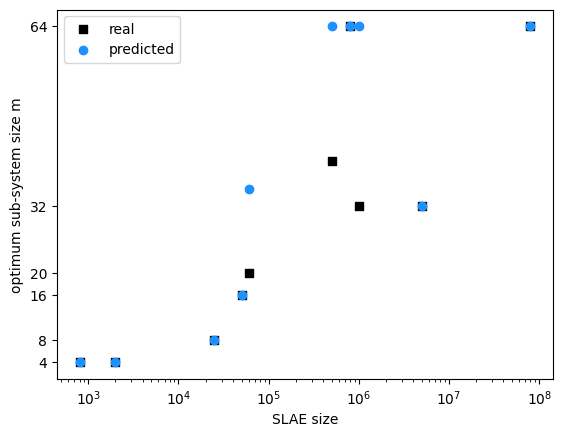}}}}%
\caption{Results from the kNN classification model for the optimum sub-system size.}
\label{fig:knn_res}
\end{figure}

The first approach gave us a 0.7 normalised accuracy score for the test set, which means that it manages to find the expected sub-system size in 7 out of 10 cases. On the other hand, using the second idea, we get a 1.0 normalised accuracy score, that is, 100\% accuracy.  Scatter plots of the achieved results can be seen in Figure~\ref{fig:knn_res}. There, the Figure to the left shows the 1.0 achieved score when using the corrected data for $m$, and the Figure to the right illustrates the 0.7 achieved score when using the original data for $m$. The incorrect predictions are shown in black. Note that in Figure~\ref{fig:knn_res} in most cases the real and the predicted data points coincide. Finally, let us investigate the null accuracy. Null accuracy is called the accuracy that could be achieved by always predicting the most frequently met sub-system size. In our case, it was found to be 0.4. Hence, the model accuracy is 1.0 with a null accuracy of 0.4. Thus, we can conclude that the 1-NN classification model we built is doing a good job. Changing from sub-optimal to optimal (bigger) size gives us a speed-up up to 1.7 times, which was achieved for SLAE size $N= 8\times10^{7}$ comparing $m = 64$ and $m = 4$.

\subsection{Comparison of the results with memory alignment considerations}
Considering preservation of memory alignment when using
CUDA streams and when partitioning the data, the following observations can be made.
Memory allocated through the CUDA Runtime API, such
as cudaMalloc, is guaranteed to be aligned to at least 256
bytes. However, this is not the case if we use offsets
which is needed when the application relies on CUDA
streams. In order to keep the memory alignment 
the sub-system size should be multiple of 32 (for FP64). This aligns well with the result (note that for SLAE size $N\geq8\times10^{5}$ the optimum sub-system size is 32 or 64). 

\section{Recursive parallel partition method}
\label{sec:rec}

As the next step, we formulate an algorithm for finding the optimum sub-system size for the recursive parallel partition method. 
In the introduction, we sketched the stages of the non-recursive partition method.
However, instead of solving the interface system in Stage 2, we can apply the partition method again. This would mean applying one recursive step. We can do this several times. See Figure~\ref{fig:recursive} for the comparison of the operations needed for the non-recursive and recursive partition methods. Note that the sizes of the boxes do not correspond to the time needed for the respective operation.  

\usetikzlibrary{decorations.pathmorphing} % noisy shapes
\usetikzlibrary{fit}					% fitting shapes to coordinates
\usetikzlibrary{backgrounds}	% drawing the background after the foreground
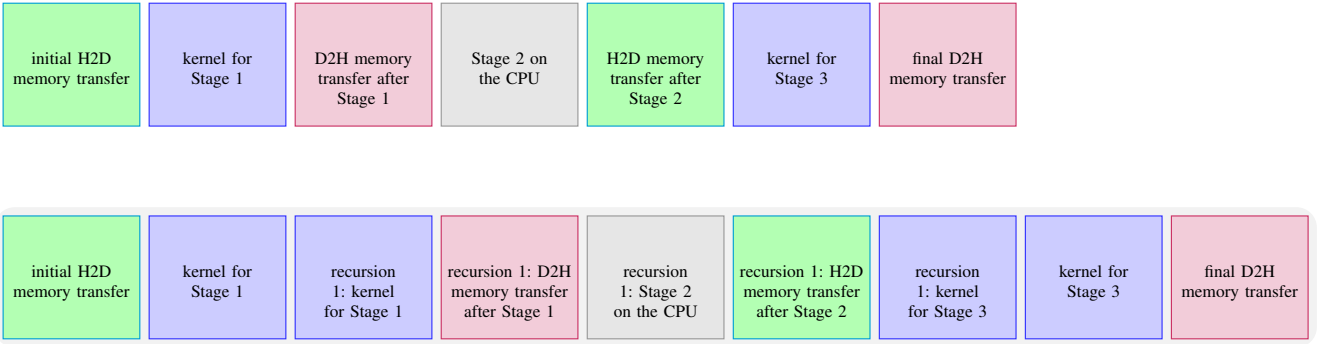
\begin{figure}[htb]
\centering
\tikzstyle{state}=[rectangle,thick,minimum size=3cm,draw=blue!80,fill=blue!20]
\tikzstyle{measurement}=[rectangle,thick,minimum size=3cm,draw=orange!80,fill=orange!25]
\tikzstyle{input}=[rectangle,thick,minimum size=3cm,draw=purple!80,fill=purple!20]
\tikzstyle{matrx}=[rectangle,thick,minimum size=3cm,draw=gray!80,fill=gray!20]
\tikzstyle{noise}=[rectangle,thick,minimum size=3cm,draw=cyan!85!black,fill=green!30!white]

\tikzstyle{state1}=[rectangle,thick,minimum width=1cm,minimum height=3cm,draw=blue!80,fill=blue!20]
\tikzstyle{measurement1}=[rectangle,thick,minimum size=1cm,draw=orange!80,fill=orange!25]
\tikzstyle{input1}=[rectangle,thick,minimum size=1cm,draw=purple!80,fill=purple!20]
\tikzstyle{matrx1}=[rectangle,thick,minimum size=1cm,draw=gray!80,fill=gray!20]
\tikzstyle{noise1}=[rectangle,thick,minimum size=1cm,draw=cyan!85!black,fill=green!30!white]

\tikzstyle{background}=[rectangle,fill=gray!10,inner sep=0.2cm,rounded corners=5mm]
\tikzstyle{backgroundPack}=[rectangle,thick,minimum width=5.5cm,minimum height=1cm,draw=purple!80,inner sep=0.2cm, rounded corners=5mm]
\tikzstyle{backgroundUnpack}=[rectangle,thick,minimum width=5.5cm,minimum height=1cm,draw=orange!80,inner sep=0.2cm,rounded corners=5mm]
\tikzstyle{backgroundInner}=[rectangle,thick,minimum width=1.5cm,minimum height=1cm,draw=blue!80,inner sep=0.2cm,rounded corners=5mm]
\tikzstyle{backgroundOuter}=[thick,draw=green!85!black,rounded corners=5mm]

\resizebox{1.0\textwidth}{!}{
\begin{tikzpicture}[>=latex,text height=0em,text depth=0.25ex,font=\large]
\matrix[row sep=0.05cm,column sep=0.2cm, ampersand replacement=\&, text width=3.1cm, align=center] {
\node (H2D0) [noise]{initial H2D memory transfer}; 	\&
\node (kernel1) [state]{kernel for Stage 1}; 	\&
\node (D2H) [input]{D2H memory transfer after Stage 1}; 	\&
\node (Thomas) [matrx]{Stage 2 on the CPU}; \&
\node (H2D1) [noise]{H2D memory transfer after Stage 2}; 	\&
\node (kernel3) [state]{kernel for Stage 3}; 	\&
\node (D2H1) [input]{final D2H memory transfer}; 	\&
\\[6em]
\node (H2D0) [noise]{initial H2D memory transfer}; 	\&
\node (kernel1) [state]{kernel for Stage 1}; 	\&
\node (kernel1) [state]{recursion 1: kernel for Stage 1}; 	\&
\node (D2H) [input]{recursion 1: D2H memory transfer after Stage 1}; 	\&
\node (Thomas) [matrx]{recursion 1: Stage 2 on the CPU}; \&
\node (H2D1) [noise]{recursion 1: H2D memory transfer after Stage 2}; 	\&
\node (kernel3) [state]{recursion 1: kernel for Stage 3}; 	\&
\node (kernel3) [state]{kernel for Stage 3}; 	\&
\node (D2H1) [input]{final D2H memory transfer}; 	\&
\\
};
\begin{pgfonlayer}{background2}
\node [background,fit=(H2D0)(D2H1)] {};
\end{pgfonlayer}{background2}

\end{tikzpicture}
}
\caption{Operations of the non-recursive partition method (top), and the recursive partition method with one recursive step (bottom).}
\label{fig:recursive}
\end{figure}

\subsection{Optimum number of recursive steps}

\begin{figure}[!htb]
\centering
{\fbox{\includegraphics[width=0.45\textwidth]{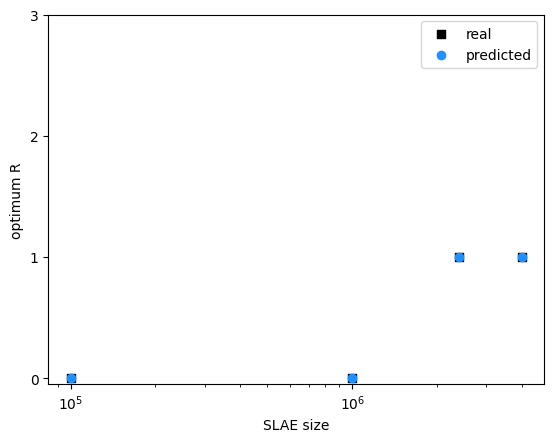}}}
\caption{Results from the kNN classification model for the optimum number of recursive steps.}
\label{fig:knn_r}
\end{figure}

Firstly, we conducted computational experiments on NVIDIA RTX A5000 to find the optimum number of recursive steps for very many different SLAE sizes. The SLAE sizes were chosen to find the cut-lines in the intervals: $10^{5}$, $1, 2, 2.2, 2.3, 2.4, 2.5, 3, 4, 4.5, 4.8, 5, 8, 8.4, 9.2, 9.6$$\times10^{6}$, $10^{7}$ and $10^{8}$. The results for 4 SLAE sizes are shown in Figure~\ref{fig:recursive_2}, while the intervals are summarized in Table~\ref{tab:opt_R}. Using this information, we built a 1-NN model for the optimum number of recursive steps. It achieves a 1.0 normalised accuracy score for the test set and a 0.5 null accuracy. Figure~\ref{fig:knn_r} shows the results. Note that in this Figure the real and the predicted data points coincide. 
\pgfplotsset{
    select row/.style={
        x filter/.code={\ifnum\coordindex=#1\else\def\pgfmathresult{}\fi}
    },
    %cycle list/Set1-9,
}

\pgfplotstableread[col sep=comma,header=true]{
A,p0,p1,p2,p3,p4
$10^{5}$,0.704061,0.8647671,0.8682630,0.9244519,0.9688411
$10^{6}$,3.176656,3.301060,3.420162,3.458871,3.760116
$10^{7}$,27.67224,26.04078,25.39645,25.07047,26.51914
$10^{8}$,271.9130,243.6584,242.6711,241.1610,243.2405
}\datatable
\begin{figure}[!htb]
\centering
\resizebox{0.75\textwidth}{!}{%
\begin{tikzpicture}
  \begin{axis}[
    ybar, 
    ymode=log,
    %cycle list/Set1-9,
    %bar shift=0pt,
    %enlarge y limits={value=.1,upper},%0.1,
    % xmin=0,
    %xtick={0,...,26},
    %xticklabels from table={\datatable}{0},
    xlabel={A},
    xtick=data,
    xticklabels from table={\datatable}{A},
    ymajorgrids = true,
    major grid style={draw=white},
    ymax=1000,
    axis x line*=bottom,
    axis y line*=right,
    y axis line style={opacity=0},
    tickwidth=1pt,
    bar width=4mm, 
    height=9cm, width=16.5cm,
    xlabel={SLAE size},
    ylabel={Time [ms]},
    x tick label style={font=\small,rotate=0},
    %nodes near coords align={horizontal},
    every axis plot post/.append style={
        draw=none,
        fill=.!65,
    },
    nodes near coords,
    nodes near coords align=vertical,
    point meta=rawy,
    %nodes near coords,
    every node near coord/.append style={
        font=\small, 
        text=black,
        rotate=90,
        anchor=west,
        /pgf/number format/precision=2,
        /pgf/number format/sci,},
        %\pgfmathprintnumber[fixed zerofill, precision=1]
        %{\pgfplotspointmeta}
        legend pos=north west
    ]

%    \pgfplotsinvokeforeach{0,...,9}{
      \addplot table [x expr=\coordindex, y=p0] {\datatable};
      \addlegendentry{Part(0)}
      \addplot table [x expr=\coordindex, y=p1] {\datatable};
      \addlegendentry{Part(1)}
      \addplot table [x expr=\coordindex, y=p2] {\datatable};
      \addlegendentry{Part(2)}
      \addplot table [x expr=\coordindex, y=p3] {\datatable};
      \addlegendentry{Part(3)}
      \addplot table [x expr=\coordindex, y=p4] {\datatable};
      \addlegendentry{Part(4)}
%    }
  \end{axis}
\end{tikzpicture}
}
\caption{Comparison between the times for the partition method with different number of recursions.}
\label{fig:recursive_2}
\end{figure}
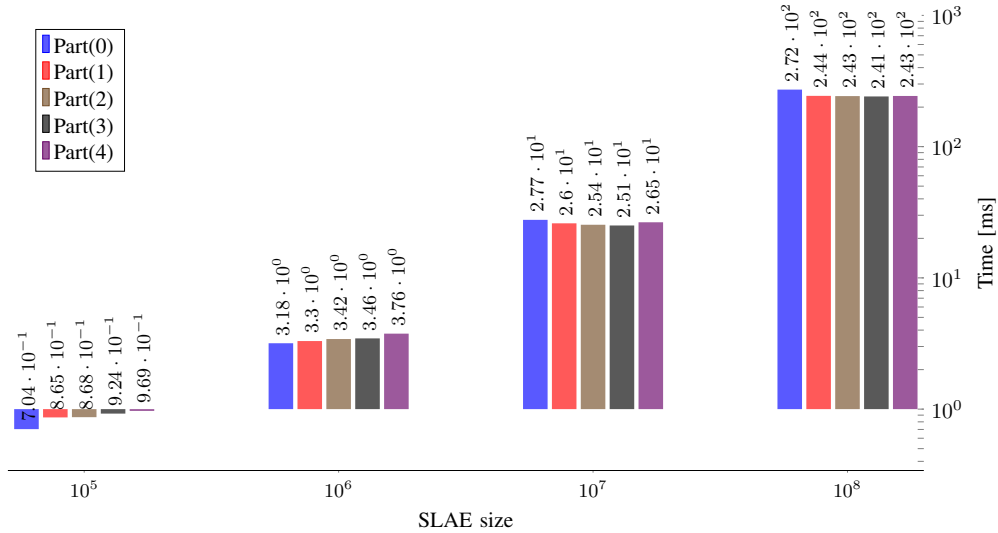
\begin{table}[!htb]
\caption{Intervals of SLAE sizes for which 4, 3, 2, 1, and 0 recursive steps lead to the smallest computational times. R denotes the number of recursive steps.}
\label{tab:opt_R}
\centering
\resizebox{0.3\textwidth}{!}{%
\begin{tabular}{|c|l|}
\hline 
R & SLAE sizes \\\hline\hline
4 & N/A \\
3 & [$10^{7}$; $10^{8}$] \\
2 & [$5\times10^{6}$; $9.6\times10^{6}$] \\
1 & [$2.3\times10^{6}$; $4.8\times10^{6}$] \\
0 & up to SLAE size $2.2\times10^{6}$ \\
\hline\hline
\end{tabular}}%}
\end{table}

\subsection{Formulation of the algorithm for the optimum \texorpdfstring{$m_{i}$}{Lg}}
The computational experiments helped us to formulate the following algorithm for choosing the sub-system sizes for R recursions:
\begin{itemize}
\item 
\underline{all recursions:} $m = $ the optimum sub-system size for the initial SLAE size (according to the already built heuristic),
\item 
\underline{if $R = 1$:} $m_{1} = $ the optimum sub-system size for the 1st interface system,\\
\underline{else:} for $m_{1}$ try small sizes like $4,5,8$ or $10$ (see the Remark below),
\item 
$m_{2}= $ the optimum sub-system size for the 2nd interface system, 
\item 
$m_{3}= $ the optimum sub-system size for the 3rd interface system.
\item
$m_{4}= $ the optimum sub-system size for the 4th interface system.
\end{itemize}
\textbf{Remark:} in 6 out of the 9 cases $m_{1} = 10$, and the difference between the worst and the best computational time when using
4, 5, 8 or 10 is negligible as a percentage of the computational
time. Hence, we can fix $m_{1}$ to 10.

When using recursive approach we managed to achieve a speed-up up to 1.17 times in comparison with the non-recursive algorithm. This was achieved for SLAE size $N=4.5\times10^{6}$. 

\section{Other experiments}
\subsection{Experiments on other NVIDIA GPU cards}
The computational experiments on NVIDIA RTX A5000 show that the heuristic for the optimum sub-system size changes slightly, and we can reuse the one found on RTX 2080 Ti with a loss of performance up to 9.44\%.
On the other hand, when running experiments on NVIDIA RTX 4080~\cite{4080}, we can reuse the heuristic found on RTX 2080 Ti with a performance loss up to 7.13\%. 
Among the many SLAE sizes that were tested, in 8 cases, applying the heuristic found on RTX 2080 Ti on RTX 4080 results in a significant performance loss (greater than 2.5\%). In 6 of these cases (given in bold in Table~\ref{tab:diff_GPU_card_fp64}) the optimum sub-system size found for RTX 4080 coincides with the optimum sub-system size found for RTX A5000.   
%On RTX 4080 in 6 out of the 8 cases (given in bold in Table~\ref{tab:diff_GPU_card_fp64}) when the heuristic (on RTX 2080 Ti) gives significant differences (above 2.5\%) between the times when using the optimum sub-system size for RTX 4080, and the time when using the heuristic size, the optimum sub-system size on RTX 4080 coincides with the optimum sub-system size on RTX A5000.  
Overall, from the results achieved on RTX A5000 and RTX 4080 we can conclude that using one and the same heuristic on these cards would lead to no performance loss. All results are summarized in Table~\ref{tab:diff_GPU_card_fp64}.
\begin{table}[!htb]
\caption{Observations on the optimum sub-system size, FP64, different GPU cards. The 6th and the 8th columns show the performance loss on A5000 and 4080, respectively, when using the heuristic that was found on 2080~Ti.}
\label{tab:diff_GPU_card_fp64}
\centering
%\resizebox{0.8\textwidth}{!}{%
{\scriptsize
\begin{tabular}{|r|r|r|r|r|r|r|r|}
\hline
$N$ & \#streams & opt $m$ & heuristic & opt $m$ & difference with & opt $m$ & difference with\\
&       & 2080 Ti & on 2080~Ti& A5000 & the 2080~Ti heuristic & 4080 & the 2080~Ti heuristic \\\hline\hline
$10^{2}$         & 1     & 4       & 4         & 4     & --   & 4    & --\\
$2\times10^{2}$  & 1     & 4       & 4         & 4     & --   & 4    & --\\
$4\times10^{2}$  & 1     & 4       & 4         & 4     & --   & 4    & --\\
$5\times10^{2}$  & 1     & 4       & 4         & 4     & --   & 4    & --\\
$8\times10^{2}$  & 1     & 4       & 4         & 4     & --   & 8    & small \\
$10^{3}$         & 1     & 4       & 4         & 4     & --   & 4    & --\\
$2\times10^{3}$  & 1     & 4       & 4         & 4     & --   & 4    & --\\
$4\times10^{3}$  & 1     & 4       & 4         & 8     & small& 8    & small \\
$4.5\times10^{3}$& 1     & 4       & 4         & 4     & --   & 4    & --\\\hline
$5\times10^{3}$  & 1     & 8       & 8         & 4     & small& 4    & small \\
$8\times10^{3}$  & 1     & 8       & 8         & 8     & --   & 4    & small \\
$10^{4}$         & 1     & 8       & 8         & 8     & --   & 8    & --\\ 
$2\times10^{4}$  & 1     & 8       & 8         & 8     & --   &16    & small \\
$2.5\times10^{4}$& 1     & 8       & 8         & 8     &      & 8    & --\\\hline
$3\times10^{4}$  & 1     &16       &16         &16     & --   &16    & --\\
$4\times10^{4}$  & 1     &16       &16         &16     & --   &16    & --\\
$5\times10^{4}$  & 1     &16       &16         &16     & --   &16    & --\\\hline
$6\times10^{4}$  & 1     &20       &20         &32     &2.65\%&40    & small \\
$7\times10^{4}$  & 1     &35       &20         &20     & --   &20    & --\\
$7.5\times10^{4}$& 1     &40       &20         &20     & --   &40    & small \\\hline
$8\times10^{4}$  & 1     &32       &32         &40     & small&32    & --\\
$10^{5}$         & 1     &40       &32         &32     & --   &32    & --\\ 
$2\times10^{5}$  & 2     &64       &32         &\textbf{64}     &6.26\%&\textbf{64}    &4.59\%\\
$4\times10^{5}$  & 3     &64       &32         &64     &3.54\%&64    & small\\
$5\times10^{5}$  & 8     &40       &32         &40     &2.38\%&40    &4.19\%\\
$8\times10^{5}$  & 8     &64       &32         &64     &6.03\%&64    &2.50\%\\
$10^{6}$         & 8     &32       &32         &\textbf{64}     &9.44\%&\textbf{64}    &7.13\%\\ 
$2\times10^{6}$  &16     &32       &32         &\textbf{64}     &8.15\%&\textbf{64}    &6.00\%\\
$4\times10^{6}$  &32     &32       &32         &\textbf{64}     &5.60\%&\textbf{64}    &6.90\%\\
$5\times10^{6}$  &32     &32       &32         &\textbf{64}     &3.65\%&\textbf{64}    &5.66\%\\
$8\times10^{6}$  &32     &64       &32         &64     &5.63\%&64    &7.09\%\\
$10^{7}$         &32     &32       &32         &\textbf{64}     &6.06\%&\textbf{64}    &6.75\%\\\hline
$2\times10^{7}$  &32     &64       &64         &64     & --   &64    & --\\
$4\times10^{7}$  &32     &64       &64         &64     & --   &64    & --\\
$5\times10^{7}$  &32     &64       &64         &64     & --   &64    & --\\
$8\times10^{7}$  &32     &64       &64         &64     & --   &64    & --\\
$10^{8}$         &32     &64       &64         &64     & --   &64    & --\\ 
\hline\hline
\end{tabular}}%}
\end{table}
%\vspace{-1em}
\subsection{Experiments with FP32 precision}
\begin{figure}[!htb]
\centering
\subfloat[Results from the kNN classification model for the optimum sub-system size using the corrected data for $m$; FP32.]{{\fbox{\includegraphics[width=0.45\textwidth]{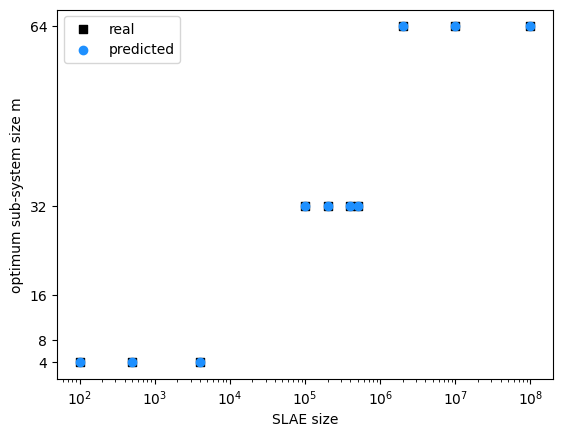}}}}
\qquad\subfloat[Results from the kNN classification model for the optimum sub-system size using the observed data for $m$; FP32.]{{\fbox{\includegraphics[width=0.45\textwidth]{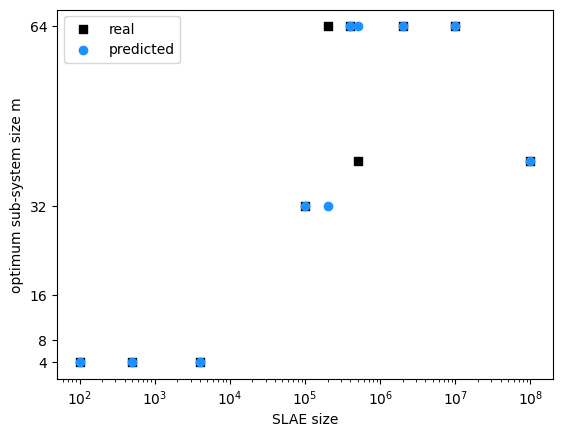}}}}%
\caption{Results from the kNN classification model for the optimum sub-system size; FP32.}
\label{fig:knn_res_fp32}
\end{figure}

Unlike the optimum CUDA stream heuristic~\cite{Veneva_2024}, here we could not observe any connection between the heuristic for FP64 and FP32. Therefore, a separate heuristic was built. The 1-NN classification model yields a 0.8 normalised accuracy score for the test set for the observed
data, a 1.0 normalised accuracy score for the test set for the corrected
data, and 0.4 for the null accuracy (see Figure~\ref{fig:knn_res_fp32}). The results are summarized in Table~\ref{tab:my_label_no_wd_fp_32}.
%\vspace{-1em}
%\vspace{-2em}
\begin{table}[!htb]
\caption{Observations on the optimum sub-system size, FP32.}
\label{tab:my_label_no_wd_fp_32}
\centering
\resizebox{0.8\textwidth}{!}{%
\begin{tabular}{|r|r|r|r|r|}
\hline
$N$ & opt $m$ & \#streams & corrected opt $m$ & time difference \\
& (1) & & (2) & (1)-(2) \\\hline\hline
$10^{2}$ & 4 & 1 & 4 & --\\ % 4: 2.877830e-01; 8: 
$2\times10^{2}$ & 4 & 1 & 4 & --\\ % 4: 2.887790e-01; 
$4\times10^{2}$ & 4 & 1 & 4 & --\\
$5\times10^{2}$ & 4 & 1 & 4 & --\\
$8\times10^{2}$ & 4 & 1 & 4 & --\\
$10^{3}$ & 4 & 1 & 4 & --\\
$2\times10^{3}$ & 4 & 1 & 4 & --\\
$4\times10^{3}$ & 4 & 1 & 4 & --\\ % 4: 3.174990e-01; 3.390150e-01
$4.5\times10^{3}$ & 4 & 1 & 4 & --\\\hline % 4: 3.515880e-01; 8: 
$5\times10^{3}$ & 8 & 1 & 8 & --\\ % 4: 3.431750e-01; 8: 3.297650e-01; 
$8\times10^{3}$ & 8 & 1 & 8 & --\\ % 4: 3.497700e-01; 8: 3.494560e-01; 16: 3.518220e-01
$10^{4}$ & 8 & 1 & 8 & --\\\hline % 4: 3.664990e-01; 8: 3.548010e-01; 
$2\times10^{4}$ & {\color{blue}16}, 8 for FP64 & 1 & 8 & \textbf{-0.016753} \\ % 4: 4.462940e-01; 8: 4.233840e-01; 16: 4.066310e-01
$2.5\times10^{4}$ & {\color{blue}20}, 8 for FP64 & 1 & 8 & -0.005121\hspace{0em} \\\hline % 4: 4.876730e-01; 8: 4.479290e-01; 20: 4.428080e-01
$3\times10^{4}$ & 16 & 1 & 16 & -- \\ % 4: 5.344810e-01; 8: 4.780540e-01; 16: 4.527360e-01; 20: 4.611280e-01
$4\times10^{4}$ & 16 & 1 & 16 & -- \\ % 4: 6.111490e-01; 8: 5.270140e-01; 16: 4.896240e-01; 
$5\times10^{4}$ & 16 & 1 & 16 & -- \\ % 4: 6.799670e-01; 8: 5.844120e-01; 16: 5.393440e-01; 20: 5.419760e-01; 32: 
$6\times10^{4}$ & {\color{blue}16}, 20 for FP64 & 1 & 16 & --\\ % 4: 7.643380e-01; 8: 6.316240e-01; 16: 5.767730e-01; 20: 5.945760e-01; 40: 
$7\times10^{4}$ & {\color{blue}16}, 35 for FP64 & 1 & 16 & --\\\hline % 16: 6.232750e-01; 20: 6.407550e-01; 35: 6.568600e-01; 40: 6.367540e-01
$7.2\times10^{4}$ & {\color{blue}32}, -- for FP64 & 1 & 32 & --\\ % 4: 8.520030e-01; 8: 7.014800e-01; 16: 6.400030e-01; 20: 6.481640e-01; 32: 6.277990e-01
%$7.5\times10^{4}$ & {\color{blue}20}, 40 for FP64 & 1 & -- & 20\\\hline % 4: 8.662940e-01; 8: 7.253099e-01; 20: 6.515020e-01; 40: 6.568160e-01
$8\times10^{4}$ & 32 & 1 & 32 & --\\ % 4: 9.105260e-01; 8: 7.496499e-01; 16: 6.849120e-01: 20: 6.901510e-01; 32: 6.786259e-01 
$10^{5}$ & {\color{blue}32}, 40 for FP64 & 1 & 32 & --\\ % 4: 1.070944e+00; 8: 8.683900e-01; 16: 7.612960e-01; 20: 7.685100e-01; 32: 7.492530e-01; 40: 7.532950e-01 
$2\times10^{5}$ & 64 & 2 & 32 & -0.052007\hspace{0em} \\ % 4: 1.831937e+00; 8: 1.439273e+00; 16: 1.291924e+00; 32: 1.225047e+00; 40: 1.236243e+00; 64: 1.173040e+00
$4\times10^{5}$ & 64 & 4 & 32 & -0.03352\hspace{0.5em} \\ % 4: 3.289651e+00; 8: 2.447554e+00; 16: 2.126383e+00; 32: 1.976410e+00; 40: 1.974458e+00; 64: 1.942890e+00
$5\times10^{5}$ & 40 & 8 & 32 & -0.03166\hspace{0.5em} \\ % 16: 2.527974e+00; 20: 2.475149e+00; 32: 2.383237e+00; 40: 2.351577e+00
$6\times10^{5}$ & {\color{blue}64}, -- for FP64 & 8 & 32 & -0.034738\hspace{0em} \\ % 16: 2.922814e+00; 20: 2.843311e+00; 32: 2.727296e+00; 40: 2.712089e+00; 64: 2.692558e+00
$7\times10^{5}$ & {\color{blue}40}, -- for FP64 & 8 & 32 & \textbf{-0.148239} \\\hline % 16: 3.304245e+00; 20: 3.212100e+00; 32: 3.186470e+00; 40: 3.038231e+00; 64: 3.257174e+00
$7.2\times10^{5}$ & {\color{blue}64}, -- for FP64 & 8 & 64 & --\\ % 16: 3.406886e+00; 20: 3.290480e+00; 32: 3.257331e+00; 40: 3.129427e+00; 64: 3.098008e+00
$8\times10^{5}$ & 64 & 8 & 64 & -- \\% 16: 3.723662e+00; 20: 3.607378e+00; 32: 3.609442e+00; 40: 3.419327e+00; 64: 3.397833e+00
$10^{6}$ & {\color{blue}64}, 32 for FP64 & 8 & 64 & -- \\ % 16: 4.528786e+00; 20: 4.505672e+00; 32: 4.458047e+00; 40: 4.326214e+00; 64: 4.172107e+00
$2\times10^{6}$ & {\color{blue}64}, 32 for FP64 & 16 & 64 & -- \\ % 16: 8.559244e+00; 20: 8.328862e+00; 32: 8.323901e+00; 40: 8.257306e+00; 64: 8.039771e+00
$4\times10^{6}$ & {\color{blue}64}, 32 for FP64 & 32 & 64 & -- \\ % 16: 1.643257e+01; 20: 1.606441e+01; 32: 1.546578e+01; 40: 1.544796e+01; 64: 1.528350e+01
$5\times10^{6}$ & {\color{blue}64}, 32 for FP64 & 32 & 64 & -- \\ % 16: 2.042002e+01; 1.977860e+01; 32: 1.904253e+01; 40: 1.879090e+01; 64: 1.864643e+01
$8\times10^{6}$ & 64 & 32 & 64 & -- \\ % 16: 3.205397e+01; 20: 3.103439e+01; 32: 2.934734e+01; 40: 2.916307e+01; 64: 2.868257e+01
$10^{7}$ & {\color{blue}64}, 32 for FP64 & 32 & 64 & --\\\hline % 16: 4.000134e+01; 20: 3.843207e+01; 32: 3.632481e+01; 40: 3.579260e+01; 64: 3.532542e+01
$2\times10^{7}$ & 64 & 32 & 64 & --\\ % 16: 7.807642e+01; 20: 7.530683e+01; 32: 7.109922e+01; 40: 6.973753e+01; 64: 6.804819e+01; 
$4\times10^{7}$ & {\color{blue}40}, 64 for FP64 & 32 & 64 & \textbf{-5.0433}\hspace{1em} \\ % 16: 1.544494e+02; 20: 1.486538e+02; 32: 1.399861e+02; 40: 1.372812e+02; 64: 1.423245e+02; 
$5\times10^{7}$ & {\color{blue}40}, 64 for FP64 & 32 & 64 & \textbf{-4.0701}\hspace{1em} \\ % 16: 1.924638e+02; 20: 1.855735e+02; 32: 1.747691e+02; 40: 1.712206e+02; 64: 1.752907e+02
$8\times10^{7}$ & {\color{blue}40}, 64 for FP64 & 32 & 64 & \textbf{-7.1318}\hspace{1em} \\ % 16: 3.069785e+02; 20: 2.952781e+02; 32: 2.832792e+02; 40: 2.723114e+02; 64: 2.794432e+02
$10^{8}$ & {\color{blue}40}, 64 for FP64 & 32 & 64 & \textbf{-11.3864}\hspace{0.75em} \\ % 16: 3.822587e+02; 20: 3.693508e+02; 32: 3.581428e+02; 40: 3.481058e+02; 64: 3.594922e+02
\hline\hline
\end{tabular}}%}
\end{table}

\section{Discussion and Conclusions}

This paper presented a heuristic for finding the optimum sub-system size for the CUDA implementation of the parallel partition algorithm by applying tools frequently used in the modern AI-focused approaches. 
Computational experiments for different SLAE sizes were conducted, and the optimum sub-system size for each of them was found empirically. Then, we used the collected data to build a 1-NN classification model for prediction of the optimum sub-system size given the SLAE size.
Changing from sub-optimal to optimal (bigger) size gave us a speed-up up to 1.7 times, which was achieved for SLAE size $N= 8\times10^{7}$ comparing $m = 64$ and $m = 4$. The heuristic for the optimum sub-system size is in good correspondence with the memory alignment requirements. The results show that the optimum sub-system size does not come when we use the largest number of CUDA blocks, although theoretically this would make the GPU busier. By comparing the predicted values with the actual data, the algorithm was deemed to be acceptably good. Next, the heuristic was expanded to work for the recursive parallel partition algorithm as well, and an algorithm for determining the optimum sub-system size for each recursive step was formulated. When using recursive approach we managed to achieve a speed-up up to 1.17 times, achieved for SLAE size $N=4.5\times10^{6}$, in comparison with the non-recursive algorithm. A 1-NN model for predicting the optimum number of recursive steps for a particular SLAE size was built. Interestingly, it turns out that solving an SLAE of any size does not get faster when using the partition method with four recursive steps. Most probably, the overhead from the additional recursion is bigger than the performance gain achieved by making smaller D2H (before Stage 2) and H2D (after Stage 2) memory transfers and solving a smaller interface system during Stage~2.

\section*{Acknowledgement}
The author would like to thank Dr.\,Alexander Ayriyan (JINR) and  Prof.\,Toshiyuki Imamura (R-CCS, RIKEN)  for their precious comments.

\IEEEpeerreviewmaketitle

\end{document}